\documentclass[prl,twocolumn,amssymb,showpacs]{revtex4}

\usepackage{graphicx}
\usepackage{bm}

\begin{document}

\title{Condensate entanglement and multigap superconductivity in
nanoscale superconductors}

\author{R. Saniz, B. Partoens, and F. M. Peeters}

\affiliation{Departement Fysica, Universiteit Antwerpen,
Groenenborgerlaan 171, B-2020 Antwerpen, Belgium}

\date{\today}

\pacs{74.20.-z, 74.78.-w, 74.81.-g}

\begin{abstract}
A Green functions approach is used to study superconductivity in
nanofilms and nanowires. We show that the superconducting
condensate results from the multimodal entanglement, or internal
Josephson coupling, of the subcondensates associated with the
manifold of
Fermi surface subparts resulting from size-quantisation.
This entanglement is of critical importance in these systems,
since without it superconductivity would
be extremely weak, if not completely negligible.
Further, the multimodal character
of the condensate generally results in multigap superconductivity,
with great quantitative consequence for the value of the critical
parameters. Our approach suggests that these are universal
characteristics of confined superconductors.

\end{abstract}

\maketitle

In recent years, superconductivity has been studied with great
interest in nanoscale systems,
such as nanofilms (NFs) \cite{guo04,eom06,qin09,zhang10} and
nanowires (NWs) \cite{tian05,zgirski07}. From the theory
point of view, there was interest in superconductivity in
NFs \cite{blatt63} long before the advent of nanoscience.
The great advances in materials synthesis
technology, however, allows today
the fabrication of high quality nanoscale
samples, in which the effects of confinement, or
quantum size effects, can be closely examined.
Thus, for instance, the
oscillations of the superconducting gap
with film thickness predicted in Ref.~\onlinecite{blatt63}
have now been observed convincingly in experiment, albeit with a
weaker amplitude than the one predicted \cite{guo04,eom06,qin09}.
Recent theoretical work has shown similar quantum size-induced
oscillations in the critical temperature, specific heat,
and critical field in NFs
\cite{chen06,shanenko07}
as well as in NWs \cite{han04,shanenko06}. Interestingly,
it is found that,
for sufficiently small confinement length (i.e., film thickness or
wire radius), the oscillations can drive the critical temperature well
above the bulk value. This was not observed in the ultrathin NFs
of Refs.~\onlinecite{guo04,eom06,qin09}, but
appears to be the case in the NFs of
Ref.~\onlinecite{court08}, and also
in NWs \cite{tian05,zgirski07}.

Here we show that confinement has far reaching consequences
regarding the character of the superconducting
condensate itself. In this regard, the
study of superconductivity in nanosystems
is of broader interest. Indeed, superfluidity also fascinates
researchers in atomic physics \cite{schunck08} and
nuclear physics \cite{pastore08}, where confinement occurs only
naturally. Superfluidity is a macroscopic quantum phenomenon,
in which the order parameter, or condensate ``wave function''
plays a central role. Thus, much effort is put in trying to
characterise it in the different systems in which superfluidity
is observed \cite{schunck08,fernandes10}.
In this work, we use a Green
functions approach to superconductivity in nanosystems,
bringing to light previously unrecognised but significant
confinement effects.
We find that the splitting of the Fermi surface in a
discrete set of
nonintersecting parts because of the size-quantisation of the energy 
levels results inevitably in a
condensate of composite nature. That is, the
condensate arises from the multimodal
entanglement, or internal Josephson coupling \cite{leggett66,hines03},
of the subcondensates associated with the
Fermi surface subparts. Further, the entanglement is crucial,
because without it the systems collapses into uncorrelated
subcondensates, in which superconductivity is dramatically weaker,
if not completely negligible.
Finally, the manifold of subcondensates
will generally result in multigap
superconductivity, with significant impact on the predicted value of
the critical temperature, compared to single-gap models.
The generality of our formalism suggests that these
characteristics are universal properties of confined
superconductors.

Consider a system of quasiparticles with a weak
attractive effective interaction
coupling only particles with
opposite spin (e.g.,
the net effect of phonon
exchange and the screened Coulomb interaction), which will eventually
couple only particles near the Fermi level (chemical potential),
$\mu$. We have the Hamiltonian
$\hat H=\hat H_0+\hat H_I-\mu\hat N$, 
where $\hat H_0$ describes the noninteracting system, $\hat N$ is
the number operator, and
$\hat H_I=
{1\over 2}\int d^3r\,d^3r'
\hat\psi_\uparrow^\dagger({\bf r})
\hat\psi_\downarrow^\dagger({\bf r}')v_{\rm eff}({\bf r},{\bf r}')
\hat\psi_\downarrow^\dagger({\bf r}')
\hat\psi_\uparrow({\bf r}).$
Consider next the Green function,
${\cal G}({\bf r}{\bf r}',\tau)=-\langle T_\tau
\hat\psi_\uparrow^\dagger({\bf r}\tau)
\hat\psi_\uparrow^{\phantom\dagger}({\bf r}'0)\rangle$.
As in BCS theory \cite{fetter71}, to solve the equation of
motion for ${\cal G}$ we introduce the Gorkov functions
${\cal F}({\bf r}{\bf r}',\tau)=-\langle T_\tau
\hat\psi_\uparrow({\bf r}\tau)
\hat\psi_\downarrow({\bf r}'0)\rangle$
and
${\cal F}^\dagger({\bf r}{\bf r}',\tau)=-\langle T_\tau
\hat\psi_\downarrow^\dagger({\bf r}\tau)
\hat\psi_\uparrow^{\phantom\dagger}({\bf r}'0)\rangle$, and
take the mean-field approximation
$\langle T_\tau
\hat\psi_\downarrow^\dagger({\bf r}_1\tau_1)
\hat\psi_\downarrow^{\phantom\dagger}({\bf r}_2\tau_2)
\hat\psi_\uparrow^{\phantom\dagger}({\bf r}_3\tau_3)
\hat\psi_\uparrow^\dagger({\bf r}_4\tau_4)\rangle
\simeq -{\cal F}({\bf r}_3{\bf r}_2,\tau_3-\tau_2)
{\cal F}^\dagger({\bf r}_1{\bf r}_4,\tau_1-\tau_4)$.
The resulting coupled equations for ${\cal G}$ and
${\cal F}^\dagger$ read, in frequency domain,
\begin{eqnarray}
&&\hspace{-10pt}
{\cal L}_p({\bf r})\tilde{\cal G}({\bf r}{\bf r}',\omega_p)
-\int d^3x\Delta({\bf r}{\bf x})
\tilde{\cal F}^\dagger({\bf x}{\bf r}',\omega_p)
=\hbar\delta({\bf r}-{\bf r}'), \nonumber \\
&&\hspace{-10pt}
{\cal L}_p({\bf r})
\tilde{\cal F}^\dagger({\bf r}{\bf r}',\omega_p)-
\int d^3x\Delta({\bf r}{\bf x})
\tilde{\cal G}({\bf x}{\bf r}',\omega_p)=0. \label{bcseqs-r} 
\end{eqnarray}
Here, ${\cal L}_p({\bf r})=i\hbar\omega_p-H_0({\bf r})+\mu$,
with $\omega_p$ is a fermionic frequency,
$\tilde{\phantom{G}}$ denoting a $\tau$-Fourier transformed
function, and
we have introduced $\Delta({\bf r}{\bf r}'')\equiv
v_{\rm eff}({\bf r},{\bf r}'')
{\cal F}({\bf r}{\bf r}'',0)$, i.e., a nonlocal pairing
potential \cite{pastore08,bruder90}.

Assume now that $H_0$ is such that its
eigenstates form an orthonormal set, i.e.,
$H_0|\nu\rangle=E_\nu|\nu\rangle$,
with $\langle\nu|\nu'\rangle=\delta_{\nu,\nu'}$.
Inserting the expansions
$\hat\psi_\sigma({\bf r}\tau)=
\sum_\nu\psi_\nu({\bf r})c_{\nu\sigma}(\tau)$ and
$\hat\psi^\dagger_\sigma({\bf r}\tau)=
\sum_\nu\psi^*_\nu({\bf r})c^\dagger_{\nu\sigma}(\tau)$
in Eqs.~(\ref{bcseqs-r}) leads to [with, e.g.,
$\tilde{\cal G}({\bf r}{\bf r}',\omega_p)=\sum_{\nu\nu'}
\psi_\nu({\bf r})\psi^*_{\nu'}({\bf r}')
\tilde{\cal G}(\nu\nu',\omega_p)$]
\begin{eqnarray}
&&\hspace{-15pt}
\left[i\hbar\omega_p-\epsilon_\nu\right]
\tilde{\cal G}(\nu\nu',\omega_p)
+\sum_\kappa\Delta(\nu\kappa)\tilde{\cal F}^\dagger(\kappa\nu',\omega_p)
=\hbar\delta_{\nu\nu'}, \nonumber \\
&&\hspace{-15pt}
\left[i\hbar\omega_p+\epsilon_\nu\right]
\tilde{\cal F}^\dagger(\nu\nu',\omega_p)+
\sum_\kappa\Delta^*(\nu\kappa)\tilde{\cal G}(\kappa\nu',\omega_p)=0. \label{gfeq}
\end{eqnarray}
Here,
$\epsilon_\nu=E_\nu-\mu$ and
$\Delta(\nu\nu')=\sum_{\kappa\kappa'}V_{\nu\nu',\kappa\kappa'}{\cal F}(\kappa\kappa',0),$
with $V_{\nu\nu',\kappa\kappa'}\equiv=-
\langle\nu\nu'|v_{\rm eff}|\kappa\kappa'\rangle$.
The formal solution of Eqs.~(\ref{gfeq}) can readily be written in
closed form. But we further assume that $v_{\rm eff}$ couples
only time-reversed states \cite{anderson59}.
If $|-\!\nu\rangle$ and $|\nu\rangle$
denote time-reversed states, we have
$
V_{\nu\nu',\kappa\kappa'}=V_{\nu\,-\!\nu,\kappa\,-\!\kappa}
\delta_{-\nu\,\nu'}\delta_{-\kappa\,\kappa'}\equiv
V_{\nu\kappa}
\delta_{-\nu\,\nu'}\delta_
{-\kappa\,\kappa'}
$
and
$\Delta(\nu\nu')=\delta_{-\nu\nu'}
\sum_{\kappa}V_{\nu\kappa}{\cal F}(-\!\kappa\,\kappa,0)\equiv\Delta(\nu)
\delta_{-\nu\nu'}$. 
As in BCS theory, it is straightforward to deduce that
$\tilde{\cal G}(\nu\nu',\omega_p)=-\delta_{\nu\nu'}
(i\omega_p+\epsilon_\nu/\hbar)/
(\omega_p^2+\xi_\nu^2/\hbar^2)$
and
$\tilde{\cal F}^\dagger(\nu\nu',\omega_p)=\delta_{-\nu\nu'}
(\Delta^*(\nu)/\hbar)/(\omega_p^2+\xi^2_\nu/\hbar^2),$
with $\xi^2_\nu=\epsilon_\nu^2+|\Delta(\nu)|^2$. Thus,
the gap values are given by the coefficients of the expansion
of $\Delta({\bf r}{\bf r}'')$ over the quasiparticle states.
Finally, noting that ${\cal F}(-\!\kappa\,\kappa,0)=\sum_p
\tilde{\cal F}(-\!\kappa\,\kappa,\omega_p)/\beta\hbar,$ we obtain
the gap equation
\begin{equation}
\Delta(\nu)=\sum_{\nu'}V_{\nu\nu'}\Delta(\nu')
{1\over{2\xi_{\nu'}}}\tanh{{\xi_{\nu'}}\over{{2k_{\rm B}T}}}.
\label{gapeq}
\end{equation}
Once $\Delta(\nu)$ is determined, the condensate wave function,
defined by
$\Psi({\bf r}{\bf r}')\equiv {\cal F}({\bf r}{\bf r}',0)$ \cite{leggett08},
and the pairing potential can be calculated from
\begin{eqnarray}
&&\hspace{-10pt}
\Psi({\bf r}{\bf r}')=\sum_\nu 
\psi_\nu({\bf r})\psi_{-\nu}({\bf r}')\Delta(\nu)
{1\over{2\xi_\nu}}\tanh{{\xi_\nu}\over{2k_{\bf B}T}}, \label{psirrp} \\
&&\hspace{-10pt}
\Delta({\bf r}{\bf r}')=\sum_{\nu}
\psi_\nu({\bf r})\psi_{-\nu}({\bf r}')\Delta(\nu). \label{deltarrp}
\end{eqnarray}

In the following we apply our formalism
to NFs and NWs.
We take parameters corresponding to Al \cite{comment0},
which is a weak coupling
superconductor (so a mean-field approach is
applicable), and is also free-electron-like.
Given that the
spatial dependence of $v_{\rm eff}$
is not really known, it is best to use phenomenologically
motivated $V_{\nu\nu'}$'s in Eq.~(\ref{gapeq})
(thereby, implicitly defining the spatial
form of $v_{\rm eff}$).
We use as reference
the BCS coupling constant for the bulk material, $V_0$,
estimated from the experimental value of $T_c$ and
$k_{\rm B}T_c=1.13\hbar\omega_De^{-1/N_0V_0}$ \cite{fetter71}.
Below, unless otherwise stated, energies are
in Rydbergs (Ry), and lengths in $a_0$. 

To model a NF we follow Ref.~\onlinecite{thompson63}. Briefly,
the system of quasiparticles is in a potential
well defined by two large planes of side $L$, a distance $d$
apart, with
$V({\boldsymbol\rho},z)=0$ for $0\leq z\leq d$, and $\infty$ otherwise.
The quasiparticles states are
$\psi_{{\bf q}n}({\boldsymbol\rho},z)=
(2/L^2d)^{1/2}e^{i{\bf q}\cdot{\boldsymbol\rho}}
\sin a_nz,$
where $a_n=\pi n/d$ and
${\bf q}=2\pi(l,m)/L$, with
$l,m\in\mathbb Z$. Thus,
$|\nu\rangle=|{\bf q}n\rangle$ and
$|-\!\nu\rangle=|-\!{\bf q}n\rangle$
are time-reversed states.
The energy levels are given by
$E_{{\bf q}n}=q^2+a_n^2$
(for simplicity, the quasiparticle mass is taken equal to
the bare electron mass).
The Fermi ``surface'' breaks into
a set of concentric circumferences of
radii $q_{\rm F}^{(n)}=
(\mu-a_n^2)^{1/2}.$
For the $V_{\nu\nu'}$ in Eq.~(\ref{gapeq}) we take a BCS-type
model \cite{fetter71},
\begin{equation}
V_{{\bf q}n,{\bf q}'n'}={{U_{nn'}}\over{L^2d}}
\theta(\epsilon_w-|\epsilon_{{\bf q}n}|)
\theta(\epsilon_w-|\epsilon_{{\bf q}'n'}|), \label{vqqp}
\end{equation}
$\epsilon_w$ defining the energy window around $\mu$
within which $v_{\rm eff}$ is effective.
We estimate the $U_{nn'}$
with a contact potential $v_{\rm eff}({\bf r}-{\bf r}')=
-V_0\delta({\bf r}-{\bf r}')$ \cite{fetter71,comment1}, hence
$U_{nn'}=V_0(1+\delta_{nn'}/2)$ \cite{thompson63}.
Note that
Eq.~(\ref{vqqp}) will lead in general to a multigap equation,
similar to the expression
of Suhl {\it et al.} \cite{suhl59}.
But because the contact interaction results in off-diagonal
$U_{nn'}$'s that are all equal, there is only one gap.
Thereafter, it is straightforward to derive
the results of previous authors (cf., e.g., Refs.
\onlinecite{shanenko07,thompson63}).

It is important to recognise, however, is that even with the
simple contact potential, this still
is a system with multiple subcondensates.
To see this, we rewrite the Hamiltonian
in second-quantised form. In close
analogy to the two band case studied by Leggett
\cite{leggett66}, one finds
$\hat H=\hat H_{\rm D}+\hat H_{\rm J}$, with
\begin{eqnarray}
&&\hat H_{\rm D}=\sum_n \Big[
\sum_{{\bf q}\sigma}\epsilon_{{\bf q}n\sigma}
c^\dagger_{{\bf q}n\sigma}c^{\phantom\dagger}_{{\bf q}n\sigma}
 \nonumber \\
&&\phantom{MMMMMM}
-U_n
\sum_{{\bf q}{\bf q}'}
c^\dagger_{{\bf q}n\uparrow}c^\dagger_{-{\bf q}n\downarrow}
c^{\phantom\dagger}_{-{\bf q}'n\downarrow}
c^{\phantom\dagger}_{{\bf q}'n\uparrow}
\Big] \nonumber \\
&&\hat H_{\rm J}=\sum_{n\neq n'}\Big[
-J\sum_{{\bf q}{\bf q}'}
c^\dagger_{{\bf q}n\uparrow}c^\dagger_{-{\bf q}n\downarrow}
c^{\phantom\dagger}_{-{\bf q}'n'\downarrow}
c^{\phantom\dagger}_{{\bf q}'n'\uparrow}
\Big].
\end{eqnarray}
Here, $\hat H_{\rm D}$ is the Hamiltonian of the
independent condensates, with $U_n=3V_0/2$, while $\hat H_{\rm J}$
represents an internal Josephson coupling \cite{leggett66},
with $J=V_0$.
\begin{figure}
\includegraphics[width=0.49\hsize]{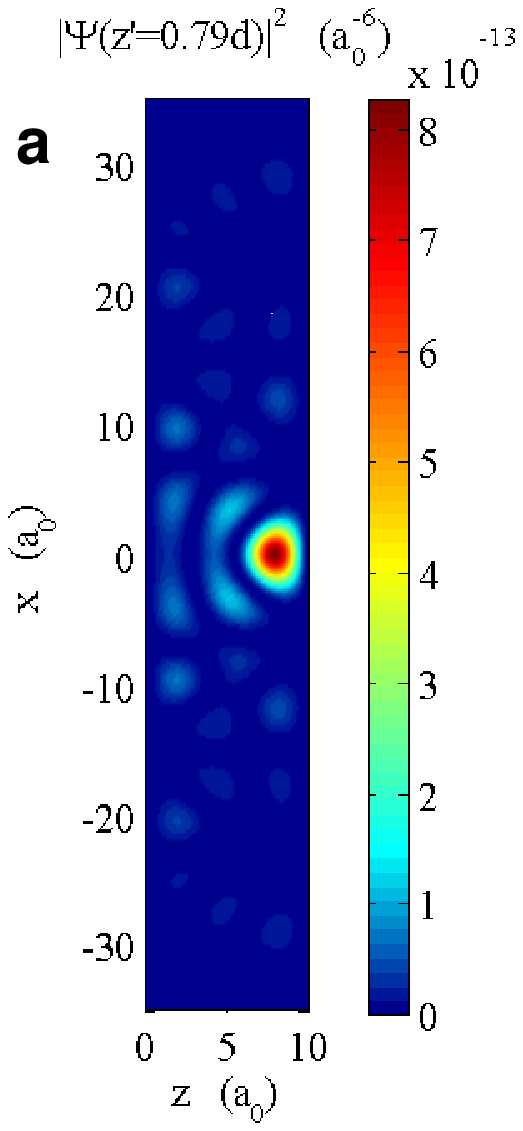}
\includegraphics[width=0.49\hsize]{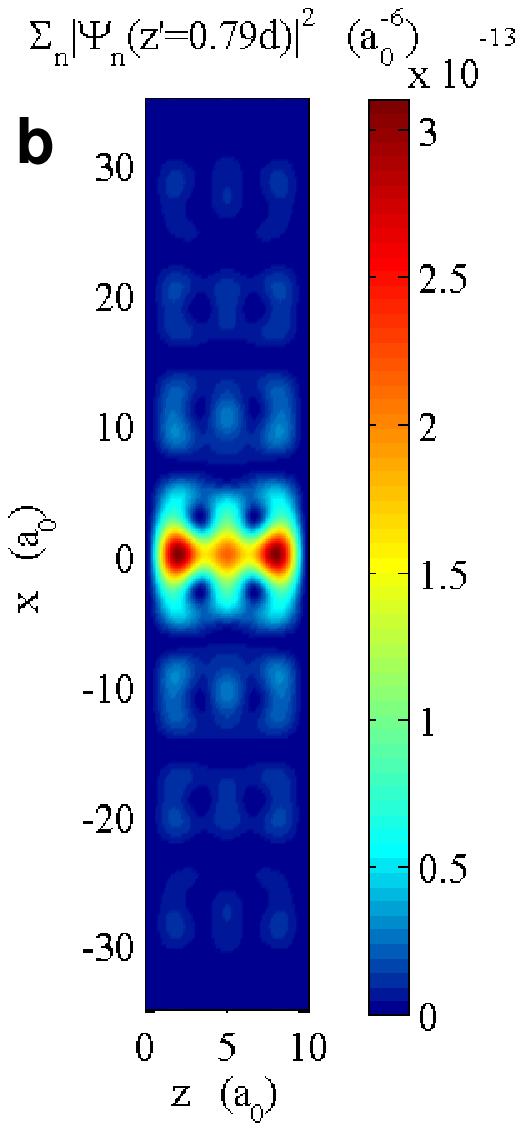}
\caption{\label{fig1} (color online)
Condensate probability density (a)
$|\sum_n\Psi_n({\boldsymbol\rho},z;z')|^2$ and
$\sum_n|\Psi_n({\boldsymbol\rho},z;z')|^2$ (b),
plotted in the $xOz$ plane, at $z'=0.79d$, for
$d=10\,a_0$ (see text).
In (b) there are no interference effects.}
\end{figure}
Thus, the condensate is
given by a multimodal entanglement of subcondensates \cite{hines03}.
The entanglement is beautifully illustrated by the resulting
interference pattern in the probability density,
$|\Psi({\bf r}{\bf r}')|^2$.
In the present case, Eq.~(\ref{psirrp}) leads to
$\Psi=\Psi({\boldsymbol\rho},z;z')=
\sum_n\Psi_n({\boldsymbol\rho},z;z')$, with
(at $T=0$ K) 
\begin{equation}
\Psi_n({\boldsymbol\rho},z;z')={\Delta\over{4\pi^2d}}
I_n({\boldsymbol\rho})
\sin a_nz
\sin a_nz', \label{xinfa}
\end{equation} 
where $I_n({\boldsymbol\rho})=\int' dq\, q
J_0(q|{\boldsymbol\rho}|)/\xi_{{\bf q}n}$ \cite{comment2}.
Given two quasiparticles of opposite spin, at $(0,z')$
and $({\boldsymbol\rho},z)$, respectively,
$\Psi({\boldsymbol\rho},z;z')$ is their pairing probability
amplitude.
In Fig.~\ref{fig1}(a), we plot
$|\sum_n\Psi_n({\boldsymbol\rho},z;z')|^2$ for
$({\boldsymbol\rho},z)$ in the $xOy$ plane,
for a $d=10\,a_0$ NF, and contrast it
to $\sum_n|\Psi_n({\boldsymbol\rho},z;z')|^2$
[cf.~Fig.~\ref{fig1}(b)], to highlight the interference effects.
To choose $z'$, we calculated first the local pair density,
$\varrho_s(z')\equiv
\int d^2\rho\,dz|\Psi({\boldsymbol\rho},z;z')|^2$.
One readily finds
$\varrho_s(z')=(\Delta/16\pi^3d)\arctan(\epsilon_w/\Delta)
\sum_n\sin^2 a_nz'$.
In Fig.~\ref{fig2}(a) we plot $\varrho_s(z')$ for three $d$
values. For $d=10\,a_0$, $\varrho_s(z')$  is
maximum at $z'=0.79d$ (note that the
number of maxima in $\varrho_s(z')$ corresponds
the number of subcondensates in the film).

Furthermore, the strength of
the $J$ coupling is of critical importance,
its magnitude largely determining the value of the
critical parameters. Indeed, for a renormalised coupling
$J=fV_0$, with $f\leq 1$, both $\Delta$ and $T_c$
fall dramatically as $f$ decreases. We illustrate
this in Fig.~\ref{fig2}(b), for $d=10$ $a_0$.
At $f=1$, the critical parameters are significantly higher than
the bulk values, namely $T_c/T_c^b=2.51$ and
$\Delta/\Delta^b=2.60$.
For $f=0$, i.e., decoupled condensates,
$\Delta/\Delta^b$ and $T_c/T_c^b$ are negligible, of
the order of
$10^{-3}.$ In contrast, in Refs.~\onlinecite{guo04,eom06,qin09},
$\Delta$ and $T_c$ are found to be a large fraction
of the bulk values, requiring
$0.5\lesssim f < 1$ in our model,
i.e., a substantial coupling.
A value $f<1$ is easily understood, since interband
pair scattering requires a minimum momentum transfer,
so has a smaller scattering phase space volume
than intraband scattering [an aspect not accounted for in
Eq.~(\ref{vqqp})]. In fact,
this may be another reason why in experiment the
critical parameters are lower than in the bulk.

\begin{figure}
\includegraphics[width=\hsize]{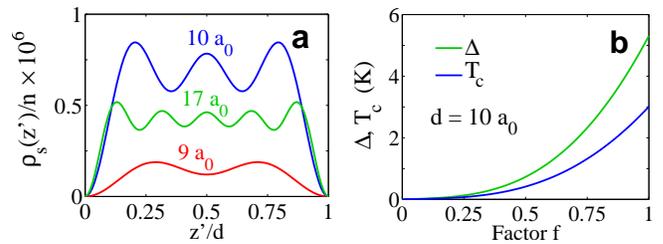}
\caption{\label{fig2} (color online)
(a) Local pair density, $\varrho_s(z')$, for three $d$ values
(in units of the bulk electron density, $n$).
The number of maxima indicate the number of subcondensates.
(b) $\Delta$ and $T_c$ change strongly with factor $f$ (see text).
At $f=0$
(decoupled condensates), $T_c\simeq 10^{-3}T_c^b$, and
$\Delta\simeq 10^{-3}\Delta^b$.
}
\end{figure}

\begin{figure}
\includegraphics[width=0.95\hsize]{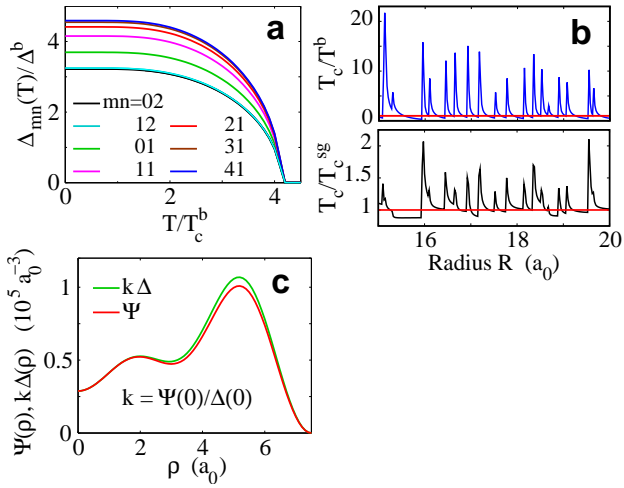}
\caption{\label{fig3} (color online)
(a) Plot of the seven $\Delta_{mn}(T)$ in a
$R=7.5\,a_0$ NW;
$T_c=4.2\,T_c^b$.
(b) {\it Upper panel}:
$T_c$ as a function of $R$. $T_c$
increases sharply when a
new band starts to be occupied. The horizontal line indicates
$T_c^b$. For large $R$, $T_c$ tends
to $T_c^b$ (not shown here). {\it Lower panel}: The
ratio of $T_c$ to the single-gap value, $T_c^{sg}$, shows
that they differ significantly.
(c) The pairing potential, $\Delta$, and order parameter,
$\Psi$, are not proportional to each other (here shown in the
${\bf r}={\bf r}'$ limit).}
\end{figure}

We now turn our attention to NWs.
The quasiparticles are now
in a cylindrical potential well of radius $R$ and length $L$:
$V(\rho,\phi,z)=0$ for $\rho\leq R$, and
$\infty$ otherwise. The quasiparticle states are \cite{han04}
$\psi_{kmn}(\rho,\phi,z)=[\pi R^2LJ_{|m|+1}^2(\eta_{mn})]^{-1/2} 
J_{|m|}(\rho\eta_{mn}/R)e^{i(kz+m\phi)},$
where $J_m$ is the $m$-th order Bessel function of the
1st kind and 
$\eta_{mn}$ is its $n$-th zero \cite{abramowitz72},
and $k=2\pi l/L$, with $l\in\mathbb Z$. Here,
$|\nu\rangle=|kmn\rangle$ and $|-\!\!\nu\rangle=
|-\!\!k\,-\!\!m\,n\rangle$ are time-reversed states.
The eigenenergies are given by $E_{kmn}=k^2+
\eta_{mn}^2/R^2$, and the
Fermi surface reduces to a discrete set
$\{-k_{\rm F}^{(mn)},k_{\rm F}^{(mn)}\}^{\phantom\dagger}_{mn}$.
The energy bands are now
1-dimensional, while they were 2-dimensional in the NFs.
This gives rise
to important quantitative differences 
between the two cases regarding
the behaviour of their properties
as a function of confining length \cite{han04,shanenko06}.
Here we focus, however, on the multigap character of
superconductivity in NWs. To see this, let us approximate
the $V_{\nu\nu'}$ in
Eq.~(\ref{gapeq}) by
\begin{equation}
V_{kmn,k'm'n'}={U_{mn,m'n'}\over{\pi^2RL}}
\theta(\epsilon_w-|\epsilon_{kmn}|)
\theta(\epsilon_w-|\epsilon_{k'm'n'}|). \label{vqqp2}
\end{equation}
To estimate the $U_{mn,m'n'}$ we again use a contact
potential. Unlike the
NF case, the off-diagonal elements are
different from each other.
This immediately results in multiple gaps, $\Delta_{mn}$.
In Fig.~\ref{fig3}(a) we plot the gap values as a function
of temperature for a $R=7.5\,a_0$ NW \cite{comment5}.
In this case
there are seven occupied bands, thus seven subcondensates.
The $\Delta_{mn}(0)$ values depend on the interplay between the
$U_{mn,m'n'}$ strengths and how far from $\mu$
are the bottoms of the bands
[recall that in 1-dimension the density of states has an
(integrable) singularity at $k=0$].
$T_c$ and the $\Delta_{mn}$ oscillate strongly
as a function $R$, rising sharply when the bottom of a newly
occupied band falls below $\mu$ as $R$ increases.
This is illustrated for $T_c$ in Fig.~\ref{fig3}(b)
(upper panel) \cite{comment6}.
Although similar to the oscillations found in
single-gap models \cite{han04,shanenko06},
the multigap character of the condensate results in
significant quantitative differences.
Indeed, Fig.~\ref{fig3}(b), lower panel, shows the plot
of the ratio of $T_c$'s obtained in the multigap and
single-gap cases (the latter, $T_c^{sg}$,
is obtained by approximating the
$U_{mn,m'n'}$ by their average value,
$\bar U_{mn,m'n'}$). We see that
$T_c^{sg}$ can be more than
100\% too low respect to the multigap value.
As one would expect, the magnitude of the $J$
coupling is just as critical here
as in NFs. Indeed, setting
$U_{mn\neq m'n'}=0$ in the gap equation results in uncorrelated
condensates, with $T_c^{(mn)}$ values largely reduced respect to
the true $T_c$. For example, in a $R=5$ $a_0$ NW,
for which $T_c=2.12\,T_c^b$,
there would be three condensates, with critical temperatures
$T_c^{(21)}\simeq 0.078\,T_c^b$,
$T_c^{(11)}\simeq 10^{-5}\,T_c^b$, and
$T_c^{(01)}\simeq 10^{-12}\,T_c^b$.

We add that, because the matrix elements decrease with
confining length, a finite $J$ 
coupling is essential to obtain the bulk values of the
critical parameters in the limit of large systems
(i.e., $R\to\infty$ in NWs and $d\to\infty$ in NFs).
Also,
as defined in Eqs.~(\ref{psirrp}) and
(\ref{deltarrp}), the pairing potential and the order parameter
are not proportional to each other (unlike in homogeneous systems
\cite{fetter71}),
even in the ${\bf r}={\bf r}'$ limit. For example,
in Fig.~\ref{fig3}(c) we compare $\Delta(\rho)$ (renormalised,
for comparison) and $\Psi(\rho)$ 
(in that limit both depend only on $\rho$) for the $R=7.5\,a_0$
wire. So our $\Delta(\rho)$ is not equivalent to the ``order
parameter'' in
other approaches \cite{han04,shanenko06}. Also, our $\Delta(\rho)$
should not be confused with the spatially varying gap seen,
e.g., in some high-$T_c$ superconductors \cite{mcelroy05}. Indeed,
in our case the gap(s) are constant throughout the system.

We thank J. Tempere for fruitful
discussions. This work was supported by FWO-Vl and the Belgian
Science Policy (IAP).


\begin{thebibliography}{99}
\bibitem{comment0} The particle density is
$r_s=2.07\,a_0$, the (bulk)
critical parameters are $T_c^b=1.2$ K and
$\Delta^b=1.7 k_B T_c^b$, and density of states
per spin ($N_0$) is obtained 
with the electron gas expression \cite{ashcroft76}; 
$\hbar\omega_D/k_B=375$ K \cite{fetter71}.
\bibitem{comment1} It is not obvious that the value of
$V_0$ is appropriate here.
But previous work shows
\cite{chen06,shanenko07,han04,shanenko06}
that it does lead to results comparable with experiment.
\bibitem{comment6} The very large $T_c/T_c^b$ ratios should
be taken with a grain of salt, since the appropriate $V_0$
value is quite uncertain.
\end{thebibliography}
\end{document}